\documentclass[a4paper, 11pt]{article}
\usepackage[pdftex, a4paper]{geometry}
\usepackage{graphicx}				
				
\usepackage{fancyhdr}
\usepackage{amsfonts}
\usepackage{dsfont}
\usepackage{mathrsfs}
\usepackage{amssymb}
\usepackage{bbm}
\usepackage{epsfig}
\usepackage{amsmath}
\usepackage{float}
\usepackage{cite}
\usepackage{blindtext}
\usepackage{algorithm} 
\usepackage{algorithmic} 
\usepackage{amsfonts}
\usepackage{mathrsfs}
\usepackage{amssymb}
\usepackage{bbm}
\usepackage{float}
\usepackage{color}
\usepackage[english]{babel}

\newtheorem{theorem}{Theorem}
\newtheorem{definition}{Definition}

\newtheorem{lemma}{Lemma}
\newtheorem{remark}{Remark}
\newtheorem{proposition}{Proposition}

\title{\bf Finite-time Convergent Gossiping}
\date{}
\author{Guodong Shi, Bo Li, Mikael  Johansson,  and Karl H. Johansson\thanks{Some preliminary results of the current paper were presented at the  21st International Symposium on Mathematical Theory of Networks and Systems (MTNS), Groningen, The Netherlands, in July 2014 \cite{MTNS}. G. Shi is with College of Engineering and Computer Science, The Australian National University, Canberra 0200, Australia. B. Li is with with Key Lab of Mathematics Mechanization, Academy of Mathematics and Systems Science, Chinese Academy of Sciences, Beijing 100190, China.   M.  Johansson and K. H. Johansson are with ACCESS Linnaeus Centre, Royal Institute of Technology, Stockholm 10044, Sweden. e-mail: guodong.shi@anu.edu.au, libo@amss.ac.cn, mikaelj@kth.se,  kallej@kth.se. }}
\begin{document}
\maketitle

\begin{abstract}
Gossip algorithms are widely used in modern distributed systems, with applications ranging from sensor networks and peer-to-peer networks to mobile vehicle networks and social networks. A tremendous research effort has been devoted to analyzing and improving the asymptotic rate of convergence for gossip algorithms. In this work we study finite-time convergence of deterministic gossiping. We show that there exists a symmetric gossip algorithm that converges in finite time if and only if the number of network nodes is a power of two, while there always exists an asymmetric gossip algorithm with  finite-time convergence, independent of the number of nodes. For $n=2^m$ nodes, we prove that a fastest convergence can be reached in $nm=n\log_2 n$ node updates via symmetric gossiping. On the other hand, under asymmetric gossip among $n=2^m+r$ nodes with $0\leq r<2^m$, it takes at least $mn+2r$ node updates for achieving finite-time convergence. It is also shown that the existence of  finite-time convergent gossiping often imposes strong structural requirements on the underlying interaction graph. Finally, we apply our results to gossip algorithms in quantum networks, where the goal is to control the state of a quantum system via pairwise interactions. We show that finite-time convergence is never possible for such systems.
\end{abstract}


\section{Introduction}\label{Sec:Introduction}
\subsection{Motivation and Related Work}
Gossip protocols have become canonical solutions in modern distributed computer systems for their simplicity and scalability \cite{Eugster2003,Jelasity2005,Shah2008}. For a network of nodes without central coordinator, gossip protocols provide an information spread mechanism in which nodes communicate pairwise along with some deterministic or randomized pair-selection algorithm \cite{Demers1987}. Formally, a gossip protocol consists of two parts \cite{Kempe2004}: an underlying  algorithm determining pairwise node interactions for point-to-point  communication, and an interaction rule built on top of the algorithm determining  the information for exchange  and the way nodes update their internal states. Gossip-based  protocols have  been adopted  to provide distributed solutions in the areas of optimization, control, signal processing, and machine learning \cite{Nedic2010,Bullo 2012,Moura2010,Rabbat2013}, {  and recently  have even been generalized to quantum information processing leading to the development of quantum gossiping algorithms \cite{Mazzarella2013a,Mazzarella2013b}.}

The convergence speed of the underlying gossip algorithm associated with a given gossip protocol, serves naturally as the primary index to the performance of the protocol. In literature, characterizations of gossip algorithm convergence focus on two basic convergence-rate metrics: information dissemination and aggregation times. The dissemination  time concerns the minimum number of steps it takes for a message starting from one node to spread across the whole network with a probability no smaller than a given level \cite{Karp 2000}. The  aggregation time concerns the minimum number of steps it takes  for nodes in the network to compute a generic function (e.g., initial values' average) to a given accuracy with a given probability \cite{Kempe2003}. These two metrics are essentially  {\em asymptotic} rates of the probability decrease for the hitting/mixing times being  smaller than the current time slot, along a Markovian process defined by the random gossiping. Various efforts have been made on bounding and optimizing these two convergence metrics \cite{Frieze 1985,Pittel1987,Karp2000, Kempe2003,Boyd2006,Boyd2004,Benezit2010,Murray2012, Doerr2012, Iutzeler2013,Liu2011}, where it has been shown that they are  determined by the pair selection mechanism and the structure of the underlying network.

{
Finite-time convergence then naturally serves as an intriguing  limit in studying the convergence properties of gossip algorithms.
In a more general domain, the possibilities and impossibilities of reaching finite-time convergence for discrete-time consensus algorithms, represented by products of stochastic matrices,  have been  systematically investigated in \cite{Finite1,Finite2,julien-automatica,julien-tac}.  These distributed algorithms have a finite computational cost, and surprisingly,  certain distributed algorithms converging in
finite time  can be  faster than any possible centralized
algorithm \cite{julien-automatica}. In this paper, we restrict our attention to deterministic gossip algorithms and study their finite-time convergence, which will,  generally speaking, provide faster information spreading than any asymptotically convergent gossip protocols.
}

\subsection{Model}  Consider a network with node set $\mathrm{V}=\{1,\dots,n\}$. Time is slotted and the value  node $i$ holds at time $k$ is denoted as $x_i(k)\in\mathbb{R}$ for $k\geq0$. The global network state is then given by $x(k)=(x_1(k) \dots x_n(k))^T$.  A symmetric deterministic gossip  algorithm~\cite{Kempe2003,Boyd2006} is defined by a sequence of node pairs $(i_{k}, j_{k})$ for $k=0, 1, \dots$ and a node state update rule
\begin{align*}
	x_{i_k}(k+1) &= \frac{x_{i_k}(k)+x_{j_k}(k)}{2};&&\\[2mm]
	x_{j_k}(k+1) &= \frac{x_{j_k}(k)+x_{i_k}(k)}{2};&&\\[2mm]
	x_{l}( k+1) &= x_{l}(k), \ \ \ l \in \mathrm{V}\backslash \{i_k, j_k\}.
\end{align*}
Note that the two selected nodes update their state to the average of the values they held prior to the interaction, while the states of all other nodes remain unchanged.

Introduce
\begin{align}\label{102}
 \mathscr{M}_n:=\Big\{ I_n-\frac{(e_i-e_j)(e_i-e_j)^T}{2}:\  i,j\in \mathrm{V} \Big\},
\end{align}
where $I_n$ is the $n$ by $n$ identity matrix, and  $e_m=(0 \dots 0\  1\  0 \dots 0)^T$ is the $n\times1$ unit vector whose $m$'th component is $1$.   We can write the class of all deterministic gossip algorithms as
\begin{align}\label{1}
 x(k+1)=P_kx(k),\ \ P_k \in \mathscr{M}_n,\  k=0,1,\dots.
 \end{align}
 Algorithm (\ref{1}) is called an asymmetric gossip algorithm if  we replace $ \mathscr{M}_n$ with \cite{Fagnani2008}
\begin{align}
 &\mathscr{M}_n^\natural:=\Big\{  I_n-\frac{(e_i-e_j)(e_i-e_j)^T}{2}:\  i,j\in \mathrm{V}\Big\}\nonumber\\
 &\ \ \ \ \ \ \ \ \ \ \ \bigcup \Big\{  I_n-\frac{e_i(e_i-e_j)^T}{2}:\  i,j\in\mathrm{V}\Big\}. \nonumber
\end{align}
In this case, it is allowed  that only one of the interacting nodes updates its state.

Let $\mathbf{1}$ denote the all-one column vector with proper dimension. We now consider the following  definition of finite-time convergence.

\medskip

\begin{definition}
Algorithm   (\ref{1})   achieves {finite-time convergence} with respect to initial value $x(0)=x^0\in \mathbb{R}^n$ if there exists an integer $T(x^0)\geq0$ such that $x(T)=P_{T-1} \cdots P_ 0 x(0)\in {\rm span}\{ \mathbf{1}\}$. \emph{Global finite-time convergence} is achieved  if such $T(x^0)$ exists for every initial value $x^0\in\mathbb{R}^n$.
\end{definition}

\medskip

Note that global finite-time convergence is equivalent to ${\rm rank} (P_{T-1} \cdots P_ 0) =1$ for some $T\geq 1$. Let $\|\cdot \|_1$ be the matrix norm defined by $\|A\|_1=\sum_{i=1}^m \sum_{j=1}^n \big|[A]_{ij}\big|$ for any $A\in \mathbb{R}^{m\times n}$ with $\big|\cdot\big|$ denoting the absolute value. We use the following definition of computational complexity of finite-time gossip algorithms:

\medskip

\begin{definition}
Let $\{P_k\}_0^\infty$ define a symmetric or asymmetric gossip algorithm.   The number of node updates up to step $t\geq 1$ is defined as
$$
\mathbf{N}_{P_{t-1}\dots P_0}:=\sum_{k=0}^{t-1} \|I_n-P_k\|_1.
$$
The computational complexity of $n$-node symmetric (asymmetric) gossiping is defined as
\begin{align*}
&\mathbf{C}_n:=\min\Big\{\mathbf{N}_{P_{t-1}\dots P_0}: {\rm rank} (P_{t-1} \cdots P_ 0) =1, \nonumber\\
 &\ \ \ P_k\in \mathscr{M}_n\ (or\ \mathscr{M}_n^\natural), k=0,\dots,t-1, t\geq 1 \Big\}
\end{align*}
whenever the above equation admits  a finite number.
\end{definition}

\subsection{Main Results}
In this paper, we obtain the following two results for symmetric and asymmetric gossip algorithms, respectively.

\medskip

\begin{theorem}\label{thmsym}
There exists a deterministic  symmetric gossip algorithm  that converges globally in finite time  if and only if there exists an integer $m\geq 0$ such that $n=2^m$. Moreover, the following statements hold.

(i) Suppose $n=2^m$. Then the fastest symmetric gossip algorithms take a total of $mn$ node updates to converge.

(ii) Suppose there exists no  integer $m\geq 0$ such that $n=2^m$. Then for almost all initial values, there exists no symmetric gossip algorithm with finite-time convergence. In fact, the initial values admitting finite-time convergent gossiping algorithms form a union of at most countably many linear spaces whose dimensions are no larger than $n-1$.
\end{theorem}

\medskip

\begin{theorem}\label{thasym}
There always exists an asymmetric gossip algorithm that converges globally in finite time. If  $n=2^m+r$ with $m\geq 0$ and  $0\leq r<2^m$, global convergence requires and can be achieved in $mn+2r$ node updates.
\end{theorem}

\medskip

The two theorems are obtained by first establishing a lower bound on the number of node updates required for reaching finite-time consensus, and then explicitly constructing gossip algorithms that converge in a finite number of steps equal to the lower bound. Although we allow every node to interact with every other node (\emph{i.e.}, we do not impose any restricted network structure on the allowed interactions), the  fastest convergent algorithms only use a subset of the edges. In fact, we  prove that for $n=4$, finite-time convergent symmetric algorithms are essentially {\it unique}. If the sequence of node pairs $(i_{k}, j_{k}),k=0,1,\dots$ is defined by an independent random process, the above deterministic finite-time convergent gossiping implies fundamental {\it robustness}  in the presence of repulsive links in light of the the Borel-Cantelli Lemma \cite{Shi2013}. Moreover, the deterministic finite-time convergent results established in the current paper can be used to derive almost sure finite-time convergence results under random gossiping models \cite{Shi-TIT}. Theorem~\ref{thmsym} and \ref{thasym} were briefly  reported in \cite{MTNS}.

\subsection{Application: Quantum Gossip Algorithms}
We apply the obtained results to recent studies on quantum gossip algorithms. In \cite{Mazzarella2013a,Mazzarella2013b}, a gossiping algorithm was introduced to quantum systems in the aim of symmetrizing  the information contained in each qubit of an $n$-qubit quantum network. Accurate operations to large-scale quantum systems play a fundamental role in quantum information processing due to the exponentially growing system dimension and the fragility of state preservation.  We reveal that any $n$-qubit quantum gossiping algorithm is equivalent to a number of decoupled  symmetric gossip algorithms, with numbers of nodes ranging from $\binom{n}{0}$ to $\binom{n}{n}$. Therefore finite-time convergence can never be achieved for any nontrivial quantum gossiping  since  $\binom{n}{0},\binom{n}{1},\dots,\binom{n}{n}$ cannot all be equal to some of power of two as long as $n\neq 2$. This result is summarized as follows.

\medskip

{

\begin{theorem}\label{propquantum}
It is impossible to reach global finite-time convergence to full symmetrization for quantum gossip algorithms over any nontrivial  (i.e., $n \neq 2$) quantum networks.
\end{theorem}
}

\medskip

 {
In Theorem \ref{propquantum}, by saying global {finite-time} convergence  to full symmetrization,  we mean that the steady symmetric  state consensus (cf., \cite{Mazzarella2013a}) is reached in some finite steps for all initial values as  proper quantum states represented by density operators. Theorem \ref{propquantum} indicates some strong impossibility of finite-time convergence to symmetric states for quantum gossiping algorithms. However,  it should be emphasized that,  the reduced states of the qubits  essentially follow the same dynamics as the classical symmetric  gossip algorithms, and therefore we can   apply Theorem \ref{thmsym} to conclude that these reduced states will converge to an agreement in finite time if and only if the number of qubits is some power of two. This point will be detailed in Section \ref{Sec:Quantum}.

  The authors of \cite{Mazzarella2013a,Mazzarella2013b} have shown some  conceptual consistency between the classical and quantum gossip algorithms from a group-theoretic perspective, and it was shown in \cite{Mazzarella2013a} that  the asymptotic convergence of quantum gossip algorithms   follows the same contraction-mapping analysis  as its classical analogue \cite{Boyd2006}. For quantum gossip algorithms, the distinction between their  finite-time convergence in reduced states and their  impossibility of reaching finite-time convergence in symmetric states  arises directly from the quantum specificities of the network.  }

\subsection{Paper Organization}
 Section~\ref{Sec:Symmetric} focuses on the analysis of symmetric gossiping. An {\em all-or-nothing} lemma is given  for general averaging algorithms for the proof of the necessity statement of Theorem~\ref{thmsym}. We also discuss the number of algorithms  reaching finite-time convergence. Section~\ref{Sec:Asymmetric} then turns to asymmetric gossip algorithms. We establish a combinatorial lemma, by which we show the necessary number of node updates. We then construct an asymmetric algorithm which converges with the given number of node updates. Section~\ref{Sec:Quantum} discusses the application of the obtained results to  quantum gossip algorithms and proves Theorem \ref{propquantum} after a brief introductory to quantum states and quantum gossip algorithms.   Finally some concluding remarks are given in Section~\ref{Sec:Conclusions}.

{
\subsection*{Notation and Terminology}
All vectors are column vectors and denoted by lower case letters. Matrices are denoted
with upper case letters. The sets of  integers, real numbers,  and complex numbers are denoted as and $\mathbb{Z}$, $\mathbb{R}$, and  $\mathbb{C}$,  respectively. Also, $\mathbb{Z}^{\geq0}$ and $\mathbb{Z}^+$ denote the sets of positive and nonnegative integers, respectively.    A finite square matrix $M\in\mathbb{R}^{n\times n}$ is called {\em stochastic} if $[M]_{ij}\geq 0$ for all $i,j$ and $\sum_j [M]_{ij}=1$ for all $i$ \cite{Latouche1999}.  A stochastic matrix $M$ is called {\em doubly stochastic} if $M^T$ is also  stochastic. Denote  $\mathsf{S}\doteq\big\{W\in\mathbb{R}^{n\times n}: \ W \mbox{ is a stochastic matrix} \big\}$ as the set of $n\times n$ stochastic matrices. Given a matrix $M\in \mathbb{C}^{m\times n}$, the vectorization of $M$, denoted by ${\rm \bf vec}(M)$, is the $mn\times 1$ column vector  $([M]_{11}, \dots,  [M]_{m1},   \dots, [M]_{1n},\dots, [M]_{mn})^T$. For all  matrices $A,B,C$ with $ABC$ well defined, it holds that ${\rm \bf vec}(ABC)=(C^T\otimes A){\rm \bf vec}(B)$ , where $\otimes$ {is} the Kronecker product \cite{Horn}.
}
\section{Symmetric Gossip Algorithms}\label{Sec:Symmetric}
In this section, we prove Theorem 1 and discuss uniqueness of finite-time symmetric gossip algorithms. The proof is structured in several steps. First, we show that the number of nodes being some power of two is necessary for the existence of a globally convergent symmetric gossip algorithm. We do so by constructively giving one particular initial value and showing that finite-time convergence cannot be achieved for this initial value. In the second step, we note that even if {\it global} finite-time convergence is impossible, there still might exist a gossip algorithm that converges in finite time  for \emph{some} initial values (say, half of $\mathbb{R}^n$).  We exclude such a possibility by showing that the initial values from which there exists a gossip algorithm converging in finite time  form a measure zero set. This is proved through an {\em all-or-nothing} property of distributed averaging algorithms.  In the third and final step of the proof, we characterize the complexity of symmetric gossiping and propose an algorithm that converges in the minimum number of steps given by the complexity bound.

\subsection{Critical Number of Nodes} We first prove the existence of the critical number of nodes by a contradiction argument.  Suppose that $n=2^{n_1} n_2$ with $n_1\geq0$ and $n_2\geq 3$ an odd integer, and suppose that there exists a finite integer $k_\ast$ and $P_0,\dots, P_{k_\ast}\in \mathscr{M}_n$ so that (\ref{1}) converges globally in $k_\ast+1$ steps.  This means that there exists a constant $c\in \mathbb{R}$ such that $x_i(k_\ast+1)=c$ for all $i\in V$.
Consider the initial value $x_1(0)=x_2(0)=\dots=x_{2^{n_1}}(0)=0$ and $x_{2^{n_1}+1}(0)=\dots=\dots,x_n(0)=2^{k_\ast+1}$. Since each element in $ \mathscr{M}_n$ is symmetric and doubly stochastic, the initial average is preserved at every iteration. Thus,
$$
c=\frac{2^{k_\ast+1}2^{n_1} (n_2-1)}{2^{n_1} n_2}=\frac{2^{k_\ast+1}(n_2-1)}{n_2}.
$$
On the other hand, it is not hard to see that $c$ is an integer for the given initial value, since pairwise averaging takes place  $k_\ast+1$ times. Consequently, we have $c=r_22^{r_1}$ with $0\leq r_1\leq k_\ast+1$ an integer and $r_2\geq 1$ an odd integer.
Therefore, we conclude that
$$
\frac{2^{k_\ast+1}(n_2-1)}{n_2}=r_22^{r_1},
$$
which implies that
\begin{align}\label{5}
2^{k_\ast+1-r_1}(n_2-1)=r_2 n_2.
\end{align}
Since the left-hand side of Eq.~\eqref{5} is an even number while the right-hand side is odd, we have reached a contradiction.  Therefore, when $n$ is not a power of two, Algorithm (\ref{1}) with symmetric updates cannot achieve global finite-time convergence no matter how $P_0,\dots, P_{k_\ast}$ are chosen.

\subsection{All-or-Nothing Lemma}

  Recall that  $\mathsf{S}$ denotes the set of $n\times n$ stochastic matrices.  Algorithm (\ref{1}) is  a special case of distributed averaging algorithms defined by products of stochastic matrices \cite{Tsitsiklis1986,Jadbabaie2003}:
\begin{align}\label{9}
 x(k+1)=W_kx(k),\ \ W_k\in \mathsf{S}.
 \end{align}
 Let $\mathsf{S}_0\subseteq \mathsf{S}$ be a subset of $\mathsf{S}$.  We define
\begin{align*}
&\mathscr{Z}_{\mathsf{S}_0}\doteq  \Big\{z\in \mathbb{R}^n: \ \exists W_0,\dots,W_s\in \mathsf{S}_0, s\geq 0\ \nonumber\\
&\ \ \ {\rm s.t.}\  W_s\cdots W_0z\in {\rm span}\{\mathbf{1}\} \Big\}.
\end{align*}

Let ${\mu}(\cdot)$ represent the standard Lebesgue measure on $\mathbb{R}^n$. We have the following lemma for the finite-time convergence of averaging algorithm (\ref{9}).

\medskip

\begin{lemma}\label{all-or-nothing}
Suppose $\mathsf{S}_0$ is a set with at most countable elements.  Then either $\mathscr{Z}_{\mathsf{S}_0}=\mathbb{R}^n$ or  ${\mu}(\mathscr{Z}_{\mathsf{S}_0})=0$. In fact, if $\mathscr{Z}_{\mathsf{S}_0}\neq\mathbb{R}^n$, then $\mathscr{Z}_{\mathsf{S}_0}$ is a union of at most countably many linear spaces whose dimensions are no larger than $n-1$.
\end{lemma}

\medskip

\begin{remark}
Lemma \ref{all-or-nothing} implies, given countably many stochastic matrices contained in a set $\mathsf{S}_0$, either for any initial value $x^0\in\mathbb{R}^n$, we can select a sequence of matrices from $\mathsf{S}_0$ so that the obtained averaging algorithm converges in finite time starting from $x^0$, or for almost all initial values, any averaging algorithm obtained by a sequence selection from $\mathsf{S}_0$ fails to converge in finite time.
\end{remark}

\begin{remark} Note that in the definition of $\mathscr{Z}_{\mathsf{S}_0}$, different initial values can correspond to different averaging algorithms.
Even if $\mathsf{S}_0$ is finite, there are still  uncountably many different averaging algorithms on the form (\ref{9}) as long as $\mathsf{S}_0$ contains at least two elements. Therefore, the proof of Lemma \ref{all-or-nothing} requires a careful structural characterization of $\mathscr{Z}_{\mathsf{S}_0}$.
\end{remark}

Noticing that $ \mathscr{M}_n$ is a finite set and utilizing  Lemma \ref{all-or-nothing}, Claim (ii) of Theorem \ref{thmsym} follows immediately. The proof of Lemma \ref{all-or-nothing} is given in Appendix A.

\subsection{Complexity}
Now let $n=2^m$ for some integer $m\geq 0$. For any given symmetric  gossip algorithm $\{P_k\}_0^{\infty}$, we define
$$
\Psi_h:=P_{h-1}\cdots P_0, \ \ h=1, 2, \dots.
$$
and let $[\Psi_h]_{ij}$ denote the $ij$-entry of $\Psi_h$. We call node $i$ {\em active} in matrix $P_k$ if the $ii$-entry of $P_k$ equals $1/2$. Define
\begin{align*}
& s_{i}(h):=\mbox{the number of $P_k$'s such that node $i$ is active in $P_k$}\\
&\mbox{ for $k=1,\dots,h-1$. }
\end{align*}
Then, the following claim holds.

\vspace{2mm}

\noindent {\it Claim.}   $[\Psi_h]_{ii} \geq {1}/{2^{s_{i}(h)}}$.

\vspace{2mm}

This claim can be easily proved using a recursive argument.

We introduce
$$
K:=\inf_k \Big\{ P_{k-1}\cdots P_0= \big({\bf{1}} {\bf{1}}^T\big)/2^m \Big\}.
$$
Invoking the claim we clearly see that $s_{i}(K)\geq m$. That is to say, when global finite-time convergence is achieved,  each node must have been active for at least $m$ times. Since only two nodes are updated in each iteration $k$,  $K$ is at least $mn/2$. It is then straightforward to see that $\mathbf{C}_n =m n$.

\subsection{A Fastest Algorithm} Let $n=2^m$. We now present  a symmetric gossip algorithm that converges globally in $nm=n\log_2 n$ node updates.
{ Such an algorithm can be easily constructed recursively: Let the $n$ nodes be  divided   into  two subsets with an equal number ($n/2$) of nodes and suppose agreement has been achieved via symmetric gossiping, respectively, for each subset of nodes. Then obviously finite-time agreement can be realized  for the $n$ nodes after  pairwise matching the nodes in the two subsets and running  a symmetric gossiping update among each of the pairs.

We remark that  essentially the same algorithm has  been proposed implicitly in Example 2.4 of \cite{FagnaniJSAC}. Moreover, such a  recursive construction is one of  the key components of
the classical  Cooley-Tukey  algorithm \cite{butterfly} for fast Fourier transform (FFT), and in fact the symmetric gossiping  algorithm that we present below  is even a special case of Cooley-Tukey arrangement for inverse discrete Fourier transform (IDFT), where the average value corresponds to  zero-frequency coefficient \cite{IDFT}. The  Cooley-Tukey  algorithm however also made use of the periodic nature of the exponential multipliers in FFT so the matching between two subsets of $n/2$ nodes needs to be carefully selected, which is not required  for reaching a simple finite-time agreement in our case. Nonetheless, for the completion of the paper we would like to make a full exposure to this algorithm.
}

Introduce the notation $M_{ij}:=I_n-{(e_i-e_j)(e_i-e_j)^T}/{2}$ and associate each node $i \in \mathrm{V}$ with the binary representation
$$
B_1\dots B_m,\;\;  B_s\in\{0,1\},\;\;  s=1,\dots,m
$$
of the value $i-1$. We denote the $k$'th digit of the binary representation of $i-1$ as $[D_k]^i$. We present the following algorithm as a matrix selection process in $ \mathscr{M}_n$:

\begin{algorithm}[htb]
\caption{Fastest Finite-time Convergence   via Symmetric Gossiping}
\label{alg:Framwork}
\algsetup{indent=2em}
\begin{algorithmic}[1]
\STATE $k\gets 0$
\FOR{$s\gets 1,\dots, m$}
    \STATE  $\mathsf{P}_s \gets \big\{M_{ij}: [D_s]^i \neq [D_s]^j,\ \mbox{and}\ [D_l]^i=[D_l]^j, l\neq s\big\}$
    \FOR{$t\gets 1, \dots, n/2$}
        \STATE $P_k \gets \mathsf{P}_s[t]$
        \STATE $k\gets k+1$
    \ENDFOR
\ENDFOR
\RETURN $P_0, \dots, P_{mn/2-1}$
\end{algorithmic}
\end{algorithm}

The algorithm proceeds in $m$ stages. In each stage $s$, a set $\mathsf{P}_s$ of all selection matrices $M_{ij}$ involving the node pairs  $\{i,j\}$,  with $[D_s]^i \neq [D_s]^j$ and $[D_l]^i=[D_l]^j, l\neq s$, is formed.
We apply the matrices  for symmetric gossiping  following the order of subsets $\mathsf{P}_1,\dots, \mathsf{P}_m$, where matrices in the same  $\mathsf{P}_s$, $s=1,\dots,m$ can be put in arbitrary order since they commute with each other (we have used $\mathsf{P}_s[t]$ to denote the $t$'th element in $\mathsf{P}_s$). It is easy to verify that after all matrices in $\mathsf{P}_s$ have been applied   there are at most $2^{m-s}$ different values left in the network for $s=1,\dots,m$.  Thus, convergence is reached after $mn=n\log_2n$ node updates.

\subsection{Discussion} Although we don't intend to discuss how the structure of the graph influences the existence and complexity of finite-time convergent gossiping, the proposed Algorithm 1 certainly only makes use of a fraction of edges, which naturally induces a graphical structure. Indeed, the construction of Algorithm 1 is inspired by ``hypercubes", whose precise definitions are given as follows:

\medskip

\begin{definition}
The Cartesian product of  a pair of graphs $\mathrm{G}_1=(\mathrm{V}_1,\mathrm{E}_1)$ and $\mathrm{G}_2=(\mathrm{V}_2,\mathrm{E}_2)$, denoted by $\mathrm{G}_1\square\mathrm{G}_2$, is defined by

 (i) the  vertex set of $\mathrm{G}_1\square\mathrm{G}_2$ is the Cartesian product of $\mathrm{V}_1$ and $\mathrm{V}_2$, denoted  $\mathrm{V}_1 \times \mathrm{V}_2$;

  (ii) there is an edge between $(v_1,v_2), (u_1,u_2)\in \mathrm{V}_1 \times \mathrm{V}_2$ in $\mathrm{G}_1\square\mathrm{G}_2$ if and only if either $v_1=u_1$ and $\{v_2,u_2\}\in \mathrm{E}_2$, or $v_2=u_2$ and $\{v_1,u_1\}\in \mathrm{E}_1$.

The  $m$-dimensional Hypercube $\mathrm{H}^{m}$ is then  defined as
\begin{align}
\mathrm{H}^{m}=\underbrace{\mathrm{K}_2\square\mathrm{K}_2 \dots \square\mathrm{K}_2}_{m\  {\rm times}}, \nonumber
\end{align}
where $\mathrm{K}_2$ is the path graph with two nodes.
\end{definition}

\medskip

\begin{figure*}[t]
\begin{center}
\includegraphics[height=1.2in]{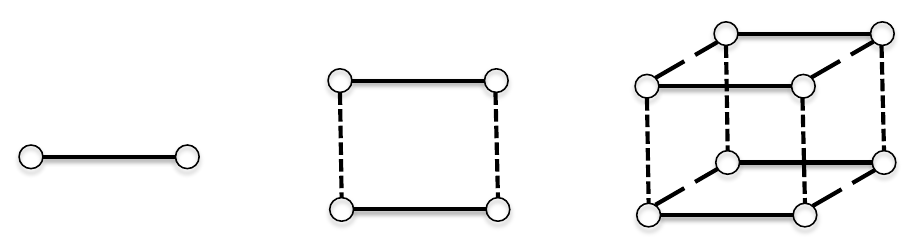}
\caption{An illustration of Algorithm 1 with $n=2$, $4$, and $8$ nodes. The edges selected in the same step are marked with the same line style. The algorithm builds hypercubes  $\mathrm{H}^{1}$,   $\mathrm{H}^{2}$, and $\mathrm{H}^{3}$. }\label{figcube}
\end{center}
\end{figure*}

In Algorithm 1, the selected edges are exactly those who form a  $
\log_2 n$-dimensional Hypercube with $n$ nodes. They are selected in the order that arises naturally from the definition of the Cartesian product (see Figure \ref{figcube}).

 We have shown that Algorithm 1 gives a fastest possible convergence. It is intriguing to ask if this algorithm is the only one that achieves finite-time consensus, or if there are (possibly many) other equally fast symmetric gossip algorithms. This turns out to be a difficult question to answer. We can, however, establish the following result indicating that for $n=4$ nodes, all finite-time convergent symmetric gossip algorithms can be reduced to an essentially unique form.

\medskip

\begin{proposition}\label{unique}
Let $n=4$. Suppose $P_{T-1} \cdots P_0 =\mathbf{1} \mathbf{1}^T/4$ with $P_k\neq P_{k+1}, k=0,\dots,T-2$ and $P_{T-2} \cdots P_0 \neq \mathbf{1} \mathbf{1}^T/4$.  Then under certain permutation of indices, we have $P_{T-1}=M_{12}$, $P_{T-2}=M_{34}$, $P_{T-3}=M_{13}$ and $P_{T-4}=M_{24}$.
\end{proposition}

\medskip

The proof of Proposition \ref{unique} is given in Appendix B.

\section{Asymmetric Gossiping}\label{Sec:Asymmetric}
In this section, we investigate  asymmetric gossiping.  We first establish a fundamental lower bound in terms of node updates for finite-time convergence,  using a combinatorial lemma. Then we construct a fastest algorithm using exactly that number of node updates.

\subsection{Complexity}
In this subsection, we first establish the least number of node updates for finite-time convergence via asymmetric gossiping. Let $n=2^m+r$ with $0\leq r<2^m$.  The following combinatorial lemma  decomposes $1$ into $n$ suitable fractions, whose proof can be found in Appendix C.

\medskip

\begin{lemma}\label{lem3}
Let $n=2^m+r$ with $0\leq r<2^m$. Introduce $\mathbb{F}\subseteq \mathbb{R}^n$ by
\begin{align*}
& \mathbb{F}=\Big\{f=(f_1,\ ...,\ f_n):\sum_{i=1}^{n}f_i=1, \nonumber\\
&\ \ \text{where}\ f_i=\frac{b_i}{2^{c_i}}, b_i,c_i \in \mathbb{Z}^{\geq 0}, b_i\ \mbox{is odd}, i=1,...,n\Big\}.
\end{align*}
For any $f\in \mathbb{F}$, we define
$$
\chi_i(f): =\inf_{d\in \mathbb{Z}} \big\{ f_i\geq \frac{1}{2^{d}}\big\},\  i=1,\dots,n.
$$  Then it holds that  $\min_{f\in \mathbb{F}}\sum_{i=1}^{n}\chi_i(f) =mn+2r$.
\end{lemma}

\medskip

Given any algorithm $\{P_k\}_0^{\infty}$,
we continue to use the notations by which we analyze the symmetric case. Recall that
$$
\Psi_h:=P_{h-1}\cdots P_0, \ \ h=1, 2, \dots.
$$
Just like the symmetric case, we define
$
s_{i}(h) $ as the number of $P_k$'s such that node $i$ is active in $P_k, k=1,\dots,h-1$
and assume the algorithm converges within $K$ steps, i.e.,  all rows of $\Psi_K$ are the same.

The following lemma follows from a simply recursive argument.

\medskip

\begin{lemma}\label{case}
For Algorithm (\ref{1}) with each $P_k\in \mathscr{M}_n^\natural$, the following always hold: (i)  $\sum\limits_{j=1}^n[\Psi_h]_{ij}=1$ for all $i$, $h$; (ii)  $[\Psi_h]_{ii} \geq {1}/{2^{s_{i}(h)}}$ for all $i$, $h$; (iii) $\sum\limits_{i=1}^n[\Psi_h]_{ij}>0$ for all $j$, $h$.
\end{lemma}

\medskip

Since all rows of $\Psi_K$ are the same, it follows that $\sum_{i=1}^{n}[\Psi_K]_{ii}=\sum_{i=1}^{n}[\Psi_K]_{1i}=1$. That is to say, $f_\ast:=([\Psi_K]_{11}\ [\Psi_K]_{22} \ ... \ [\Psi_K]_{nn})$ is an element of the set $\mathbb{F}$ defined in Lemma \ref{lem3}. Furthermore, by Lemma \ref{case}. (ii), $[\Psi_K]_{ii} \geq {1}/{2^{s_{i}(K)}}$. According to the definition of $\chi_i$ in Lemma \ref{lem3}, $s_i(K)\geq \chi_i(f_\ast)$. Therefore,
$\sum_{i=1}^{n}s_i(K) \geq \sum_{i=1}^{n} \chi_i(f_\ast) \geq mn+2r$ based on Lemma \ref{lem3}, i.e., the number of node updates is at least $nm+2r$ for reaching convergence.


\subsection{Existence}
We now construct  an algorithm that when node states converge to the same value, only $nm+2r$ node updates  have been taken.
Denote $M^*_{ij}:=I_n-{e_i(e_i-e_j)^T}/{2}$.

Again, we relabel the nodes in a binary system. We  use the binary number
$$
B_1\dots B_{m+1}, \ \ B_s\in\{0,1\}, s=1,\dots,m+1
$$
to mark node $i$ if $B_1\dots B_{m+1}=i-1$ as a binary number. We denote the $k$'th digit of $i-1$ in this binary system as $[D_k]^i$ for $k=1,\dots,m+1$ and $i=1,\dots, n$. We present the following algorithm.

\begin{algorithm}[htb]
\caption{Fastest Finite-time Convergence via  Symmetric/Asymmetric Gossiping}
\label{alg:Framwork}
\algsetup{indent=2em}
\begin{algorithmic}[1]
\STATE $k\gets 0$
\STATE $\mathsf{P}_1\gets \big\{M_{ij}: [D_1]^i \neq [D_1]^j,\ \mbox{and}\ [D_l]^i=[D_l]^j, l=2, 3,..., m+1\big\}$
\FOR{$t\gets 1, \dots, r$}
        \STATE $P_k \gets \mathsf{P}_1[t]$
        \STATE $k\gets k+1$
    \ENDFOR
\FOR{$s\gets 2,\dots, m+1$}
    \STATE  $\mathsf{P}_s^\sharp\gets \big\{M^*_{ij}: [D_1]^i=1,\ [D_1]^j=0, [D_s]^i\neq [D_s]^j, \ \mbox{and}\ [D_l]^i=[D_l]^j, l\neq s, l=2, 3,...,m+1 \big\}$
    \FOR{$t\gets 1, \dots, r$}
        \STATE $P_k \gets \mathsf{P}_s^\sharp[t]$
        \STATE $k\gets k+1$
    \ENDFOR
    \STATE $\mathsf{P}_s\gets \big\{M_{ij}: [D_1]^i=[D_1]^j=0, [D_s]^i \neq [D_s]^j, \ \mbox{and}\ [D_l]^i=[D_l]^j, l\neq s, l=2,3,..., m+1\big\}$
    \FOR{$t\gets 1, \dots, 2^{m-1}$}
        \STATE $P_k \gets \mathsf{P}_s[t]$
        \STATE $k\gets k+1$
    \ENDFOR
\ENDFOR
\RETURN $P_0, \dots, P_{(m+1)r+m2^{m-1}-1}$
\end{algorithmic}
\end{algorithm}

\begin{figure*}[t]
\begin{center}
\includegraphics[height=1.2in]{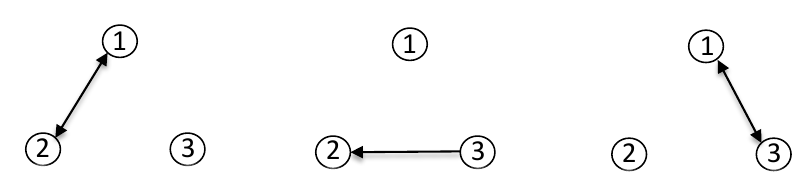}
\caption{An illustration of Algorithm 2 for three nodes. Each directed arc represents selected node pairs and only the head nodes update their states. Using three steps and five node updates, the three nodes reach the same state. } \label{fig2}
\end{center}
\end{figure*}

\begin{figure*}[t]
\begin{center}
\includegraphics[height=2.2in]{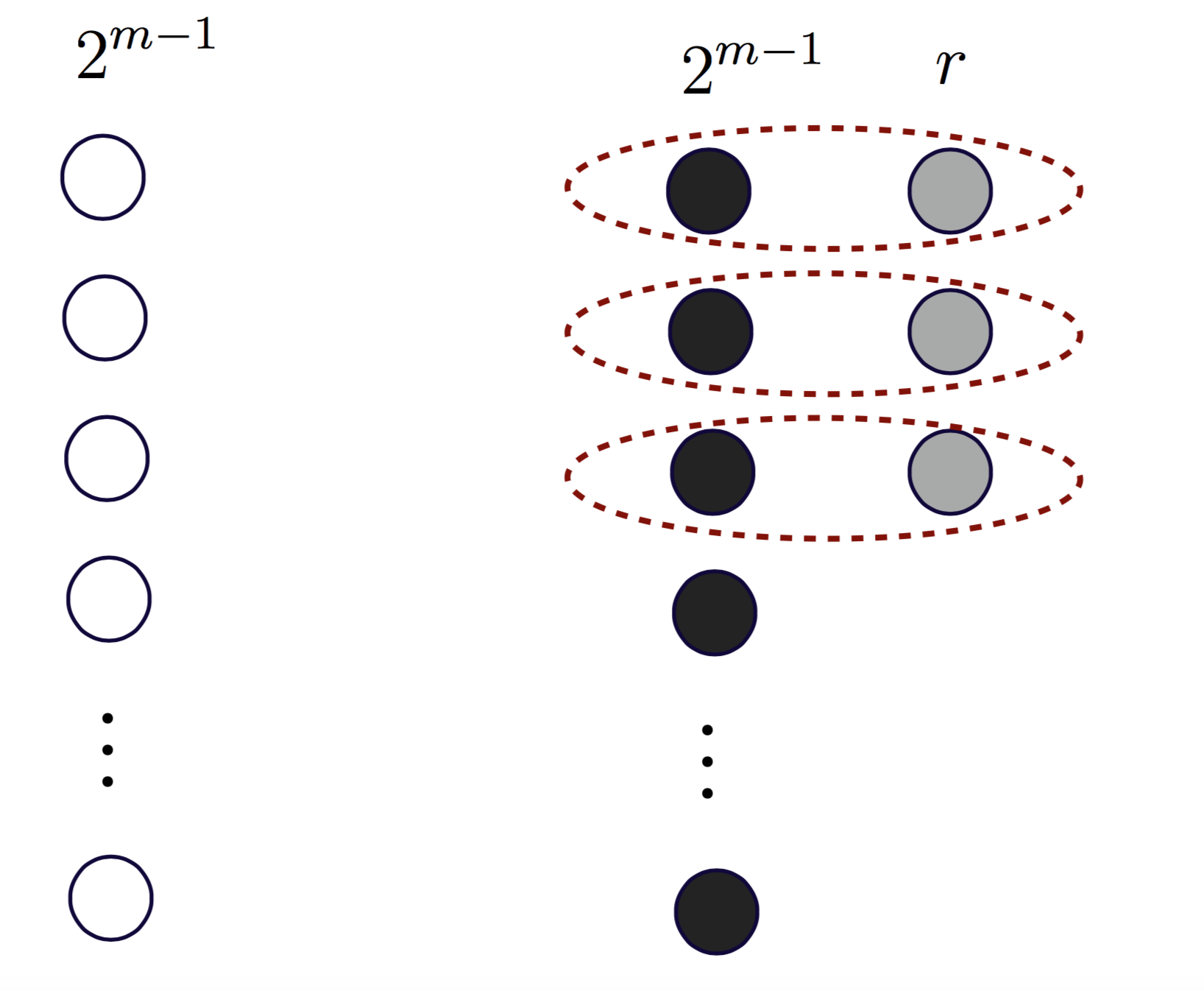}
\caption{{An illustration of Algorithm 2 for $n=2^m+r$ nodes. Divide the $2^m$ nodes into two subgroups with $2^{m-1}$ nodes in each of the group. Tie the rest of $r$ nodes pairwise with another $r$ nodes in one of the group. Then apply the above three-node arrangement to the pair of nodes with another node selected from the remaining group so that Algorithm 1 can be repeated.  Finite-time convergence is therefore achieved and it turns out this is the fastest algorithm in terms of number of node updates.}   } \label{fig5}
\end{center}
\end{figure*}

Algorithm 2 selects a sequential subsets of matrices in $\mathscr{M}^\natural$, indexed by $\mathsf{P}_1,\ \mathsf{P}_2^\sharp,\ \mathsf{P}_2,\ \mathsf{P}_3^\sharp, \ \mathsf{P}_3$, $\dots, \ \mathsf{P}_{m+1}^\sharp,\ \mathsf{P}_{m+1}$.  Matrices in the same subset can be put in arbitrary orders since they commute with each other. Matrices in $\mathsf{P}_s,s=1,\dots,m+1$ are symmetric, while matrices in $\mathsf{P}_s^\sharp,s=2,\dots,m+1$ are asymmetric. It is straightforward  to verify that after all matrices in $\mathsf{P}_s$ have been applied,   at most $2^{m-s+1}$ different value  remain  in the network. The number of node updates in Algorithm 2 can be easily calculated to be exactly $mn+2r$.

An illustration of Algorithm 2 for three nodes is shown in Figure \ref{fig2}. Note that after the first step  Node $1$ and Node $2$ hold the same value (say, $a$) and Node $3$ holds a maybe different one (say, $b$),  while the three nodes eventually agree on $({a}+{b})/{2}$ after the next  two steps. Therefore, after the first step Nodes $1$ and $2$ can be viewed as have been tied together as {\it one} node  which carries out a symmetric update with Node $3$.

Algorithm 2 is constructed based on the above intuition for three nodes. For $n=2^m+r$ nodes with distinct values, matrices in $\mathsf{P}_1$ carry out $r$ pairs of symmetric averaging and leave only $2^m$ different values. In this way $2r$ nodes are grouped into $r$ virtual nodes and then the $2^m$ different values reach finite-time convergence as in Algorithm 1 with the help of asymmetric updates (cf., Figure \ref{fig5}).

{
\begin{remark}
The  Cooley-Tukey  FFT algorithm, initially designed for a data set with a size $n=2^m$ (known as the radix-$2$ factorization) \cite{butterfly}, was later developed for general factorization forms of $n$ \cite{IDFT,FFTbook}. Such generalizations mainly used the periodicity in the exponential FFT coefficients  and generate  exact results of the FFT. This is significantly different from the idea and construction of  Algorithm 2, where it is not the exact average, but an approximate value,  is achieved. This sacrifice is anyhow inevitable if finite-time convergence is required, as suggested by the impossibility part of Theorem \ref{thmsym}.
\end{remark}
}

\begin{remark}
The rank-one matrix limit of Algorithm 2 under proper  permutation of indices can be written as $\mathbf{1} \beta ^T$, where $\beta=(\beta_1\ \beta_2,\ ...\ \beta_n)^T$ is given by
$$
\beta_i=
\begin{cases}
\frac{1}{2^{m}}, &\ i=1,\ 2,\ ...,\ n-2r;\\
\frac{1}{2^{m+1}}, &\ i=n-2r+1,\ n-2r+2,\ ...,\ n.
\end{cases}
$$
In contrast to the convergence limit of $\mathbf{1} \mathbf{1}^T/n$ under symmetric update, it can be simply computed that
$$
\Big\|\beta- \mathbf{1}/n\Big\|=\sqrt{\frac{2^m-r}{2^m+r}}\times \frac{\sqrt{2r}}{2^{m+1}}<\frac{1}{\sqrt{2^{m+1}}}<\frac{1}{\sqrt{n}},
$$
which goes to zero as the network size tends to infinity.
\end{remark}

\begin{remark}
Algorithm 2 is realized using  $r+m(n+r)/2$ matrices (and therefore $r+m(n+r)/2$ time steps) from the set $\mathscr{M}_n^\natural$. We can however find examples of $n$ and alternative algorithms that reach finite-time convergence using less than $r+m(n+r)/2$ matrices. This indicates that finding asymmetric  gossip algorithms reaching  convergence using the least time steps can be a quite different problem compared to finding algorithms using a least number of node updates.
\end{remark}

\section{Application: Quantum Gossip Algorithms}\label{Sec:Quantum}
In this section, we discuss {an} application of the obtained results to quantum gossip algorithms \cite{Mazzarella2013a,Mazzarella2013b}.

{
\subsection{Quantum Mechanics Preliminaries:  Notation and Terminology }
Information processing over quantum mechanical systems is the foundation of quantum communication and quantum computation, where fundamental challenges arise from quantum mechanics~\cite{Nielsen}. In this subsection, we give a brief introduction to quantum system states and we refer the readers  to \cite{Nielsen} for a comprehensive treatment.

\subsubsection{Quantum State Space and the Dirac Notion}
  The state space associated with any isolated quantum system is a complex vector space with inner product, {i.e.}, a Hilbert space $\mathcal{H}$. The system is completely described by its state vector, which is a unit vector in the system's state space and often denoted by $|\psi\rangle\in\mathcal{H}$  (known as the Dirac notion). The state space of a composite quantum system is the tensor product of the state space of each component system, e.g., two quantum systems with state spaces $\mathcal{H}_A$ and $\mathcal{H}_B$, respectively, form a composite system with state space $\mathcal{H}_A \otimes \mathcal{H}_B$, where $\otimes$ stands for tensor product. If the two quantum systems are isolated respectively with states  $|\psi_A\rangle\in\mathcal{H}_A$ and $|\psi_B\rangle\in\mathcal{H}_B$, the composite system admits a state $|\psi_A\rangle\otimes |\psi_B\rangle$.

\subsubsection{Density Operators}
For an open quantum system, its state can also be described by a positive (i.e., positive semi-definite) Hermitian density operator  $\rho$ satisfying $\text{tr}(\rho)=1$. A quantum state $|\psi\rangle\in\mathcal{H}$, induces a linear operator, denoted $|\psi\rangle\langle\psi |$, by
\begin{align*}
  |\psi\rangle\langle\psi | \Big(|x\rangle\Big)=  \Big(|\psi \rangle, |x\rangle\Big)  |\psi\rangle
  \end{align*}
  with $\Big(\cdot,\cdot\Big)$ being the inner product\footnote{Under Dirac notion this inner product is written as $\Big(|\psi \rangle, |x\rangle\Big)=\langle\psi |x\rangle$, where $\langle\psi |$ is the dual vector of $|\psi \rangle$.} equipped by the Hilbert space $\mathcal{H}$. Then $\rho=|\psi\rangle\langle\psi |$ defines the corresponding density operator. Density operators provide a convenient description of {\it ensembles of pure state}: If a quantum system is in state $|\psi_i \rangle$ with probability $p_i$ where $\sum_i p_i=1$, its density operator is
  \begin{align*}
  \rho=\sum_i p_i   |\psi_i\rangle\langle\psi_i |.
  \end{align*}
Any  positive and Hermitian operator  with trace one defines a proper density operator describing certain quantum state, and vice versa.

\subsubsection{Qubit Network and Swapping Operators}
The $2$-dimensional Hilbert space that forms the state-space of the most basic quantum systems is called a {\it qubit} (short for quantum bit). Let $\mathcal{H}$ be a qubit system, i.e., a two-dimensional Hilbert space. Consider a quantum network as the composite quantum system of $n$ qubits in the set $\mathrm{V}=\{1,\dots,n\}$, whose state space is within the Hilbert space $\mathcal{H}^{\otimes n}=\mathcal{H}\otimes \dots \otimes \mathcal{H}$.  The {\it swapping operator} between qubits $i$ and $j$, denoted as $U_{ij}$,  is defined by
\begin{align*}
& {U_{ij}} \big(q_{1}\otimes \dots \otimes q_{i}\otimes \dots \otimes q_j\otimes \dots \otimes q_n\big)\nonumber\\
&= q_{1}\otimes \dots \otimes q_{j}\otimes \dots \otimes q_i\otimes \dots \otimes q_n,
\end{align*}
for all $q_i\in \mathcal{H}, i=1,\dots,n$. In other words, the swapping operator $U_{ij}$ switches the information held on qubits $i$ and $j$ without changing the states of other qubits. The set of all swapping operators over the $n$-{qubit} network is denoted {by} $\mathfrak{U}:=\big\{U_{ij}:i,j\in\mathrm{V}\big\}$.

\subsubsection{Partial Trace}
Let $\mathcal{H}_A$ and $\mathcal{H}_B$ be the state spaces of  two quantum systems $A$ and $B$, respectively. Their composite system is described as a density operator $\rho^{AB}$. Let $\mathfrak{L}_A$, $\mathfrak{L}_B$, and  $\mathfrak{L}_{AB}$ be the spaces of (linear) operators over  $\mathcal{H}_A$, $\mathcal{H}_B$, and  $\mathcal{H}_A\otimes\mathcal{H}_B$, respectively.   Then the partial trace over system $B$, denoted by ${\rm Tr}_{\mathcal{H}_B}$, is an operator mapping from $\mathfrak{L}_{AB}$ to $\mathfrak{L}_{A}$ defined by
$$
{\rm Tr}_{\mathcal{H}_B}\Big(|p_A\rangle  \langle q_A| \otimes  |p_B\rangle  \langle q_B| \Big)= |p_A\rangle  \langle q_A|  {\rm Tr} \Big(  |p_B\rangle  \langle q_B| \Big)
$$
for  all $|p_A\rangle, |q_A \rangle\in \mathcal{H}_A, |p_B\rangle, |q_B\rangle \in \mathcal{H}_B$.
The reduced density operator (state) for system $A$, when the composite system is in the state  $\rho^{AB}$, is defined as $\rho^A= {\rm Tr}_{\mathcal{H}_B}(\rho^{AB})$. The physical interpretation of $\rho^A$ is that $\rho^A$ holds the full information of system $A$ in $\rho^{AB}$.

}

\subsection{Quantum Gossip Algorithms}

{Introduce a notion of time indexed by $k=0, 1, \dots$, and let $\rho(k)$ denote the density operator  of the considered $n$-qubit network at time $k$. The quantum gossip algorithm introduced in~\cite{Mazzarella2013a,Mazzarella2013b} can then be written as}
\begin{align}\label{quantum}
\rho(k+1)=\frac{1}{2} \rho(k)+ \frac{1}{2} S(k) \rho(k) S(k)^\dag,\ \
\end{align}
{where $ S(k)\in \mathfrak{U},\ k=0,1,\dots$ and $S^{\dag}$ is the conjugate transpose of the operator $S$.}

It has been shown in \cite{Mazzarella2013a,Mazzarella2013b} that under quite general (randomized or deterministic) conditions {on} the {swapping sequence}, Algorithm (\ref{quantum}) {converges} asymptotically to the  symmetric state
$$
\frac{1}{n!} \sum_{\pi \in \mathbf{P}} U_\pi \rho(0) U_\pi^\dag,
$$
where $\mathbf{P}$ is the permutation group over $\mathrm{V}$, and   $U_\pi$ is the  unitary operator over $\mathcal{H}^{\otimes n}$ defined  by
$$
U_\pi \big(q_1 \otimes \dots \otimes q_n\big)= q_{\pi(1)}\otimes \dots \otimes q_{\pi(n)},\  q_i\in \mathcal{H}, i=1,\dots,n.
$$
for any $\pi\in \mathbf{P}$.

 In the remainder of this section, we establish the proof of Theorem \ref{propquantum}. We first establish a relationship between the quantum gossip algorithm and its classical analogue. Then the conclusion follows directly from the critical node number condition and the ``all-or-nothing" lemma that we have derived earlier.
\subsection{Quantum  vs. Classical Gossiping}

For {ease} of presentation we identify the linear operators $\rho$, $U_{ij}$, and $U_\pi$ as their matrix representations in $\mathbb{C}^{2^n\times 2^n}$ under the standard computational basis of $\mathcal{H}^{\otimes n}$ in the rest of discussions. Under vectorization,   Algorithm (\ref{quantum}) can be rewritten  into the following vector form:
\begin{align}\label{vec}
{\rm \bf vec}\big(\rho(k+1)\big)=\frac{1}{2}\Big(I+ {S(k)}\otimes {S(k)}^\dag \Big) {\rm \bf vec}\big(\rho(k)\big),\ \
\end{align}
where $ {S(k)}\in \mathfrak{U},\ k=0,1,\dots$ and $I$ is the $4^n$ by $4^n$ identity matrix.

Associated with  any swapping operator $U_{ij}\in \mathfrak{U}$, we naturally define a quantum  graph, $\mathrm{G}=(\mathrm{V}, \mathrm{E})$, where $\mathrm{E}:=\big\{ \{i,j\}\big\}$ is the quantum edge set containing {only the} edge $\{i,j\}$. Since ${S(k)}\in \mathfrak{U}$ for all $k$, {each $S(k)$ can be associated with a  path graph}  $\mathrm{G}_k=(\mathrm{V}, \mathrm{E}_k)$ where $\mathrm{E}_k$ contains only one edge corresponding to the node pair in $S(k)$. {It is straightforward to verify that under the computational basis, each $S(k)$ is real, symmetric, and stochastic in $\mathbb{C}^{2^n\times 2^n}$. We further introduce $T(k):=\big(I+ S(k)\otimes S(k)^\dag \big)/2$ and make the following definition.}

\medskip

\begin{definition}
The induced graph of $T(k)$, denoted $\mathcal{G}_k=(\mathcal{V}, \mathcal{E}_k)$, {has} $\mathcal{V}=\{1,\dots,4^n\}$, and $\{m,v\}\in \mathcal{E}_k$ if only if $[T(k)]_{mv}\neq 0$ for all $m\neq v\in \mathcal{V}$.
\end{definition}

\medskip

\begin{remark}\label{remark-Tk}
Based on the matrix expression of swapping operators, it is straightforward to verify that all the nonzero  off-diagonal entries of $T(k)$ are exactly $1/2$. Since $T(k)$ is a stochastic matrix with positive diagonal entries (either $1$ or $1/2$), it means that for every row of $T(k)$ containing one nonzero (i.e., $1/2$) off-diagonal entry, its diagonal entry must be $1/2$ and the nonzero off-diagonal entry is unique. In other words, $T(k)$ carries out disjoint pairwise averaging. Consequently, $T(k)$ can be written as some finite product of  commuting  matrices within the set $ \mathscr{M}_{4^n}$. Equivalently, we can identify $T(k)\in  \mathscr{M}_{4^n}$ so that Algorithm (\ref{vec}) defines an algorithm on the form of (\ref{1}).
\end{remark}

The following lemma establishes a relationship between the two graphs $\mathrm{G}_k=(\mathrm{V}, \mathrm{E}_k)$ and  $\mathcal{G}_k=(\mathcal{V}, \mathcal{E}_k)$. The proof can be found in Appendix D.

\medskip

\begin{lemma}\label{lempartition}
For $\mathrm{G}_k=(\mathrm{V}, \mathrm{E}_k),k\in \mathbb{Z}_{\geq0}$ associated with Algorithm (\ref{quantum}), the graph $$
\bigcup_{k\geq 0}\mathcal{G}_k:= \Big(\mathcal{V}, \bigcup_{k\geq 0}\mathcal{E}_k \Big)
$$
has at least $\tau_0:={\rm dim} \big\{{\rm \bf vec}(z):  \frac{1}{n!} \sum_{\pi \in \mathbf{P}} U_\pi z U_\pi  =z\big\}$ components. This minimum number of {components in} $\bigcup_{k\geq 0}\mathcal{G}_k$ is {obtained} whenever  $\bigcup_{k\geq 0}\mathrm{G}_k:= \Big(\mathrm{V}, \bigcup_{k\geq 0}\mathrm{E}_k \Big)$ is a connected graph.
\end{lemma}

\medskip

{From now on,  without loss of generality, we} assume that   $\bigcup_{k\geq 0}\mathrm{G}_k:= \Big(\mathrm{V}, \bigcup_{k\geq 0}\mathrm{E}_k \Big)$ is connected since otherwise global convergence (asymptotic or finite time) is {obviously} impossible for Algorithm~(\ref{quantum}).  In light of Lemma~\ref{lempartition}, $\bigcup_{k\geq 0}\mathcal{G}_k$ then has $\tau_0$ connected components. There is a permutation of {the elements of $\mathcal{V}$ with associated permutation matrix} ${P^\ast}\in \mathbb{R}^{4^n \times 4^n}$ such that  Eq. (\ref{vec}) can be written as
\begin{align}\label{100}
z(k+1)= {P^\ast} T_k {P^\ast}^{-1} z(k),
\end{align}
where $z(k):= {P^\ast}{\rm \bf vec}\big(\rho(k)\big)$,  and ${P^\ast} T_k {P^\ast}^{-1}$ is {block diagonal}
$$
{P^\ast} T_k {P^\ast}^{-1}={\rm diag} \big(P_\ast^{(1)}(k),\dots,P_\ast ^{(\tau_0)}(k)\big).
$$
Here the dimension of $P_\ast^{(\varrho)}(k)$ is time-invariant and consistent with the size of the $\varrho$'th component of $\bigcup_{k\geq 0}\mathcal{G}_k$ for $\varrho=1,\dots,\tau_0$.  {Furthermore,} each $ P_\ast^{(\varrho)}(k)$ is  a symmetric gossiping matrix in the form of (\ref{102}) with a proper dimension (cf., Remark \ref{remark-Tk}). In other words, (\ref{100}) defines $\tau_0$  classical symmetric  gossip algorithms that are completely decoupled:
\begin{align}\label{101}
z^{(\varrho)}(k+1)=  P_\ast^{(\varrho)}(k) z^{(\varrho)}(k),\ \ \varrho=1,\dots,\tau_0.
\end{align}

\subsection{The Connected Components}
In this subsection, we further explore the structure of the $\tau_0$ components in $\bigcup_{k\geq 0}\mathcal{G}_k$.

We denote by $|0\rangle$ and $|1\rangle$ the standard  computational basis of $\mathcal{H}$, where $|\cdot \rangle$ represents a unit vector in $\mathcal{H}$ known as the Dirac notion \cite{Nielsen}. Let $|q_1\rangle\otimes \dots \otimes |q_n\rangle \in \mathcal{H}^{\otimes n}$ be denoted as $|q_1 \dots q_n\rangle$ for simplicity. The following is a basis of $\mathcal{H}^{\otimes n}$:
$$
\Big\{|q_1 \dots q_n\rangle: q_i\in\{0,1\}, i=1,\dots,n \Big\}.
$$
We use the notion \cite{Nielsen}
$$
|q_1 \dots q_n\rangle \langle p_1 \dots p_n|:\mathcal{H}^{\otimes n} \mapsto \mathcal{H}^{\otimes n}
$$
to denote a linear operator over $\mathcal{H}^{\otimes n}$ in that
$$
\Big( |q_1 \dots q_n\rangle \langle p_1 \dots p_n|\Big) \xi= \Big \langle  |p_1 \dots p_n\rangle, \xi \Big\rangle |q_1 \dots q_n\rangle, \ \
$$
for all  $\xi\in \mathcal{H}^{\otimes n}$, where $\big \langle \cdot, \cdot \big\rangle$ is the inner product equipped by the Hilbert space $\mathcal{H}^{\otimes n}$. We further obtain a basis for all linear operators over $\mathcal{H}^{\otimes n}$:
$$
\mathfrak{B}:=\Big\{|q_1 \dots q_n\rangle \langle p_1 \dots p_n|:  q_i, p_i\in\{0,1\}, i=1,\dots,n \Big\}.
$$

Recall that $\mathbf{P}$ denotes the permutation group over $\mathrm{V}$, in which each element $\pi$  defines a rearrangement  of indices in $\mathrm{V}$. {In particular,  we let $\pi_{ij}$ be the permutation swapping indices $i$ and $j$ with all others  unchanged.}   Associated with any $\pi\in \mathbf{P}$, we define an operator $\mathcal{F}_{\pi}$ over $\mathcal{H}^{\otimes n}\times \mathcal{H}^{\otimes n}$ by
\begin{align*}
\mathcal{F}_{\pi} \Big(|q_1 \dots q_n\rangle \langle p_1 \dots p_n| \Big)=|q_{\pi(1)} \dots q_{\pi(n)}\rangle \langle p_{\pi(1)} \dots p_{\pi(n)}|
\end{align*}
for all $ |q_1 \dots q_n\rangle \langle p_1 \dots p_n|\in \mathfrak{B}$. {Letting $\pi(k)$ be the permutation corresponding to $S(k)\in\mathfrak{U}$,} Algorithm (\ref{quantum}) can be written as
\begin{align}\label{103}
\rho(k+1)=\frac{1}{2} \rho(k)+ \frac{1}{2} \mathcal{F}_{\pi(k)} \big( \rho(k) \big).
\end{align}

Note that $\mathfrak{B}$ is a basis for the space of all linear operators over $\mathcal{H}^{\otimes n}$. Thus, it is clear from (\ref{103}) that under the basis $\mathfrak{B}$,   $\rho(k)$ is a matrix in  $\mathbb{C}^{2^n \times 2^n}$ such that $|q_1 \dots q_n\rangle \langle p_1 \dots p_n|\in \mathfrak{B}$ corresponds to an entry of $\rho(k)$, i.e., a node in $\mathcal{V}$. Furthermore,  since by our assumption $\bigcup_{k\geq 0}\mathrm{G}_k$ is connected, all the swapping permutations in $\{S(k)\}_{k\geq 0}$ form a generating subset of $\mathbf{P}$. Therefore, identifying each element $\mathfrak{B}$ to its corresponding node $\mathcal{V}$, we now see that
$$
\aleph_{|q_1 \dots q_n\rangle \langle p_1 \dots p_n|}:=\Big\{ |q_{\pi(1)} \dots q_{\pi(n)}\rangle \langle p_{\pi(1)} \dots p_{\pi(n)}|, \pi\in \mathbf{P} \Big\}
$$
is the set of nodes that are reachable from $|q_1 \dots q_n\rangle \langle p_1 \dots p_n|$ in the graph $\bigcup_{k\geq 0}\mathcal{G}_k$. In other words, for any given $|q_1 \dots q_n\rangle \langle p_1 \dots p_n| \in \mathfrak{B} \cong \mathcal{V}$, $\aleph_{|q_1 \dots q_n\rangle \langle p_1 \dots p_n|}$ defines a node subset as a connected component $\bigcup_{k\geq 0}\mathcal{G}_k$. From Lemma \ref{lempartition}, there are a total of $\tau_0$ such different $\aleph_{|q_1 \dots q_n\rangle \langle p_1 \dots p_n|}$.
\subsection{Proof of Theorem \ref{propquantum}}
In this subsection, we complete the proof of Theorem \ref{propquantum}. We proceed in {three} steps.

\noindent {Step 1.} We first consider the following set of node subsets of $\mathcal{V}$,  each of which forms one of  $\bigcup_{k\geq 0}\mathcal{G}_k$'s connected components:
$$
\mathfrak{A}^\sharp:= \Big\{\aleph_{|q_1 \dots q_n\rangle \langle p_1 \dots p_n|}:\ q_1=\dots=q_n=0, p_i \in \{0,1\} \Big\}.
$$
It is straightforward to see that fixing $|0 \dots 0\rangle \langle p_1 \dots p_n|$, we have
\begin{align*}
&\aleph_{|0 \dots 0\rangle \langle p_1 \dots p_n|}=\Big\{ |0 \dots 0\rangle \langle z_1 \dots z_n|: \ z_t\in \{0,1\}, \\
 &\ \ \ t=1,\dots,n, \ \mbox{and}\ \sum_{t=1}^n z_t = \sum_{t=1}^n p_t\Big\}.
\end{align*}
Therefore, there are $n+1$ different element in $\mathfrak{A}^\sharp$, and the number of nodes in each element ranges in
$$
\Bigg \{ \binom{n}{0},\ \binom{n}{1},\ \dots,\ \binom{n}{n} \Bigg \}.
$$
We can easily verify that for any $n\neq 2$,  at least one of the above combinatorial numbers is not some power of two. From its equivalent form (\ref{101}), we conclude from Theorem \ref{thmsym} that Algorithm~ (\ref{vec}) fails to reach finite-time convergence for all ${ \rm \bf vec} (\rho(0)) \in \mathbb{R}^{4^n}$.

\noindent{Step 2.} Next, we show that  Algorithm~(\ref{vec}) fails to reach finite-time convergence for all Hermitian matrices $\rho(0) \in \mathbb{C}^{2^n \times 2^n}$. This point is immediately  clear noticing the following two facts: (i) each state-transition matrix $T(k)=\big(I+ S(k)\otimes S(k) \big)/2$ is real so that the real and imaginary parts of $\rho(k), k\geq 0$ define two separate  algorithms in the form of (\ref{vec}) with different initial values; (ii) for any ${ \rm \bf vec} (\rho(0)) \in \mathbb{R}^{4^n}$, we can construct a Hermitian matrix $\rho^* \in \mathbb{C}^{2^n \times 2^n}$ such that ${ \rm \bf vec} (\rho(0))={\rm Re} \big({ \rm \bf vec} (\rho^*)\big)$.

\noindent{Step 3.} In this step, we finally conclude the proof making use of the ``all-or-nothing" property established in Lemma \ref{all-or-nothing}. Consider the following set
\begin{align*}
&\mathcal{I}_o:=\Big\{ { \rm \bf vec}\big({\rm Re} (\rho^\ast)\big):  \rho^\ast \in \mathbb{C}^{2^n\times 2^n},\ \rho^\ast\ \mbox{is Hermitian,}\\
&\mbox{positive semi-definite, and}\  {\rm Tr}(\rho^\ast)=1\Big\}.
\end{align*}

We treat the condition ${\rm Tr}(\rho^\ast)=1$ under the basis  $\mathfrak{B}$, i.e., we index each entry of $\rho^\ast$ by $|q_1 \dots q_n \rangle \langle p_1 \dots p_n|\in \mathfrak{B}$. Then ${\rm Tr}(\rho^\ast)=1$ is equivalent to that
\begin{align}\label{104}
\sum_{p_i \in \{0,1\}} [\rho^\ast]_{|p_1 \dots p_n \rangle \langle p_1 \dots p_n|} =1.
\end{align}
Clearly (\ref{104}) defines an $(n-1)$-dimensional subspace in $\mathbb{R}^{4^n}$. However, we see that the $2^n$ elements
$$
|p_1 \dots p_n \rangle \langle p_1 \dots p_n|,\ p_i \in \{0,1\}
$$
are within $n+1$ different connected components in  $\bigcup_{k\geq 0}\mathcal{G}_k$ (again, we have used that $\mathfrak{B}\cong \mathcal{V}$). We know from (\ref{101}) that different connected components have completely decoupled dynamics, which gives the freedom that  each $[\rho^\ast]_{|p_1 \dots p_n \rangle \langle p_1 \dots p_n|}$ can take value from $[0, |\aleph_{|p_1 \dots p_n \rangle \langle p_1 \dots p_n|}|^{-1})$ without violating (\ref{104}). Here again $|\aleph_{|p_1 \dots p_n \rangle \langle p_1 \dots p_n|}|$ represents the cardinality of $\aleph_{|p_1 \dots p_n \rangle \langle p_1 \dots p_n|}$.

Noticing also that the positive semi-definite Hermitian matrices form a convex cone, we can finally conclude that the set of values $\mathcal{I}_o$, restricted to the nodes of the $\varrho$'th component of $\bigcup_{k\geq 0}\mathcal{G}_k$, $\varrho=1,\dots,\tau_0$, can never be a countable union of at most $(N_\varrho-1)$-dimensional subspaces, where $N_\varrho$ represents the number of nodes in that component. Making use of  Lemma~\ref{all-or-nothing}, we conclude that Algorithm~(\ref{vec}) fails to reach finite-time convergence for all ${ \rm \bf vec} (\rho(0)) \in \mathcal{I}_o$. Equivalently, we have proved that  Algorithm (\ref{quantum}) fails to reach global finite-time convergence for all initial density operators. This concludes the proof of Theorem \ref{propquantum}.

{
\subsection{Further Discussion: Finite-time Convergence in Reduced States}
In this subsection, we further investigate the evolution of the reduced states of the qubits along the algorithm (\ref{quantum}). We denote by $$
\rho^m(k):= {\rm Tr}_{\otimes_{j\neq m} \mathcal{H}_j } \big(\rho(k)\big)
$$
 the reduced state of qubit $m$  at time $k$ for each $m=1,\dots,n$, where $\otimes_{j\neq m} \mathcal{H}_j$ stands for the remaining  $n-1$ qubits' space $\otimes_{j\neq m} \mathcal{H}_j$ and ${\rm Tr}_{\otimes_{j\neq m} \mathcal{H}_j }$ is the partial trace. Note that $\rho^m(k)$ contains  the information that qubit $k$ holds in the composite network state $\rho(k)$. Taking partial trace, ${\rm Tr}_{\otimes_{j\neq r} \mathcal{H}_j }$, for $r=1,\dots,n$, for the left and right hands of the algorithm (\ref{quantum}), respectively,  yields
\begin{align} \begin{cases}
\rho^m(k+1) = \big(\rho^m(k)+ \rho^s(k)\big)/2,  & \mbox{if $\{m,s\}\in\mathrm{E}_k$};\\[2mm]
\rho^s(k+1) = \big(\rho^m(k)+ \rho^s(k)\big)/2,  & \mbox{if $\{m,s\}\in\mathrm{E}_k$};\\[2mm]
	\rho^l(k+1) = \rho^l(k), & \mbox{otherwise}.
\end{cases}
\end{align}
This shows that, despite that  each $	\rho^m(k)$ is formally a density operator (i.e., a trace-one,  Hermitian matrix in  $\mathbb{C}^{2\times 2}$), their evolution is exactly the same as the  classical symmetric gossiping algorithms. We can therefore apply Theorem \ref{thmsym} to each entry of the 	$\rho^m(k)$ and conclude that

\medskip

\begin{proposition}\label{propreduced}
Following the quantum gossiping algorithm (\ref{quantum}), the reduced states of the qubits  converge globally to an agreement in finite time, i.e., there exists $T>0$ such that
 $$
 \rho^m(T)=\sum_{j=1}^n \rho^j(0)/n
 $$
 for all $\rho(0)$, if and only the number of qubits $n$ is some power of two.
\end{proposition}

\medskip

The distinction between  the statements in Theorem \ref{propquantum} and Proposition \ref{propreduced} is due to the failure of finite-time aggregation for the information beyond the reduced states in the entire quantum network state, which defines      the quantum specificities of the network.
}
\section{Conclusions}\label{Sec:Conclusions}
We proved  that there exists a symmetric gossip algorithm that converges in finite time if and only if the number of network nodes is a power of two, and for $n=2^m$ nodes,  a fastest finite-time convergence can be reached in $mn$ node updates via symmetric gossiping. We also proved that there always exists a globally finite-time convergent gossip algorithm for any number of nodes with asymmetric updates, and for $n=2^m+r$ nodes with $0\leq r<2^m$, it requires $mn+2r$ node updates for achieving a finite-time convergence. Applying the results to quantum  gossip algorithms in quantum networks, we showed  that finite-time convergence is never possible for any nontrivial quantum networks. The results add to the  fundamental understanding of  gossiping algorithms. Future challenges lie in characterizing how the complexity of finite-time convergent gossiping relates to the structure of the underlying interaction graph, and how to construct  finite-time convergent  algorithms in a distributed manner.

\medskip

\medskip
\section*{Appendix}

\subsection*{A. Proof of Lemma \ref{all-or-nothing}}

Define a function $\delta(M)$ of a matrix $M=[m_{ij}]\in \mathbb{R}^{n\times n}$ by (cf. \cite{Hajnal1958})
\begin{equation}
\delta (M)\doteq\max_j \max_{\alpha, \beta}|m_{\alpha j}-m_{\beta j}|.
\end{equation}

Given an averaging algorithm (\ref{9}) defined by $\{W_k\}_0^\infty$ with $W_k\in \mathsf{S}_0,k\geq0$. Suppose there exists an initial value $x^0\in\mathbb{R}^n$ for which $\{W_k\}_0^\infty$ fails to achieve finite-time convergence. Then  obviously $\delta(W_s\cdots W_0)>0$ for all $s\geq0$.

\vspace{2mm}
\noindent {\it Claim.} ${\rm rank}(W_s\cdots W_0)\geq2,\ s\geq0$.

Let $W_s\cdots W_0=(\omega_1 \dots \omega_n)^T$ with $\omega_i\in\mathbb{R}^n$. Since $\delta(W_s\cdots W_0)>0$, there must be two  rows in $W_s\cdots W_0$ that are not equal. Say,    $\omega_1\neq \omega_2$. Note that $W_s\cdots W_0$ is a stochastic matrix because any product of stochastic matrices is still a stochastic matrix. Thus, $\omega_i\neq 0$ for all $i=1,\dots,n$. On the other hand, if $\omega_1=c\omega_2$ for some scalar $c$, we have $1=\omega_1^T \mathbf{1}=c\omega_2^T \mathbf{1}=c$, which is impossible because $\omega_1\neq \omega_2$. Therefore, we conclude that ${\rm rank}(W_s\cdots W_0)\geq {\rm rank}({\rm span} \{\omega_1,\omega_2\})\geq2$.  The claim holds.

\vspace{2mm}
Suppose there exists some $y\in \mathbb{R}^n$ such that $y\notin \mathscr{Z}_{\mathsf{S}_0}$. We see from the claim that the dimension of  ${\rm ker}(W_s\cdots W_0)$ is at most $n-2$ for all $s\geq0$ and $W_0,\dots,W_s\in \mathsf{S}_0$.

 Now  for $s=0,1,\dots$, introduce
\begin{align*}
& \Theta_s\doteq\big\{x  \in\mathbb{R}^n:\  \exists W_0 ,\dots,W_s\in \mathsf{S}_0, \nonumber\\
&\ \ \ {\rm s.t.} \ W_s \cdots W_0  x  \in {\rm span}\{\mathbf{1}\}\big\}.
\end{align*}
Then $\Theta_s$ indicates the initial values  from which convergence is reached in $s+1$ steps. For any fixed $W_0,\dots,W_s\in\mathsf{S}_0$, we define
$$
\Upsilon_{W_s\dots W_0} \doteq \big\{z\in\mathbb{R}^n:\  \ W_s \cdots W_0  z\in {\rm span}\{\mathbf{1}\}\big\}.
$$
 Clearly $\Upsilon_{W_s\dots W_0}$ is a linear space. It is straightforward to see that $\Theta_s=\bigcup_{W_s\dots W_0\in \mathsf{S}_0} \Upsilon_{W_s\dots W_0}$, and therefore
 $$
  \mathscr{Z}_{\mathsf{S}_0}=\bigcup_{s=0}^\infty \Theta_s=\bigcup_{s=0}^\infty \bigcup_{W_s,\dots, W_0\in \mathsf{S}_0} \Upsilon_{W_s\dots W_0}.
  $$

Noticing that $z\in \Upsilon_{W_s\dots W_0}$ implies $\big(z-W_s\cdots W_0z \big)\in {\rm ker}(W_s \cdots W_0 )$, we define a linear mapping
\begin{align}
& f:\ \ \ \Upsilon_{W_s\dots W_0}\  \longmapsto\  {\rm ker}(W_s\cdots W_0)\times{\rm span}\{\mathbf{1}\}\nonumber\\
& \ \ \ \ {\rm s.t.}\ \ \ f(z)=\big(z-W_s\cdots W_0z,W_s\cdots W_0z\big)
\end{align}
 Suppose $z_1,z_2\in \Upsilon_{W_s\dots W_0}$ with $z_1\neq z_2$. It is straightforward to see that either $W_s\cdots W_0z_1=W_s\cdots W_0z_2$ or $W_s\cdots W_0z_1\neq W_s\cdots W_0z_2$ implies $f(z_1)\neq f(z_2)$. Hence, $f$ is  injective. Therefore, noting that ${\rm ker}(W_s\cdots W_0)$ is a linear space with dimension at most $n-2$, we have ${\rm dim}(\Upsilon_{W_s\dots W_0})\leq n-1$, and thus ${\mu}(\Upsilon_{W_s\dots W_0})=0$. Consequently, we conclude that
\begin{align}
{\mu}(\Theta_s)&= {\mu} \Big( \bigcup_{W_0,\dots,W_s\in \mathsf{S}_0} \Upsilon_{W_s\dots W_0}\Big)\nonumber\\
&\leq \sum_{W_0,\dots,W_s\in \mathsf{S}_0}{\mu}\big(\Upsilon_{W_s\dots W_0}\big)\nonumber\\
&=0 \nonumber
\end{align}
because any finite power set  $\mathsf{S}_0\times \dots \times \mathsf{S}_0$ is still a countable set as long as $\mathsf{S}_0$ is countable. This immediately leads to
$$
{\mu}(\mathscr{Z}_{\mathsf{S}_0})={\mu}\Big(\bigcup_{s=0}^\infty\Theta_s\Big)\leq \sum_{s=0}^\infty {\mu}(\Theta_s)=0.
$$

Additionally, since every $\Theta_s$ is a union of at most countably many linear spaces, each of dimension no more than $n-1$, $\mathscr{Z}_{\mathsf{S}_0}$ is also a union of countably many linear spaces with dimension no more than $n-1$. The desired conclusion thus follows.

\subsection*{B. Proof of Proposition \ref{unique}}
Without loss of generality, we assume that for any $k$, $P_k\neq P_{k+1}$. Given $\{P_k\}_0^\infty$, recall that $\Psi_h= P_{h-1}\cdots P_0$. We define  $[{r}_i]^h$ as the $i$'th row vector of $\Psi_h$. We continue to define
$$
\mathscr{C}(\Psi_h):= \Big| \big\{[{r}_i]^h: i=1,2,3,4\big\} \Big|
$$
as the number of different rows of $\Psi_h$. The following lemma holds.
\begin{lemma}\label{no3points}
There is no $h$ such that  the following hold simultaneously: i) $\mathscr{C}(\Psi_h)=2$; ii) there are three different elements $a$, $b$ and $c$ from $\{1,\ 2,\ 3,\ 4\}$ satisfying
$$
[{r}_a]^h=[{r}_b]^h=[{r}_c]^h.
$$
\end{lemma}
{\it Proof.} We investigate two cases.

\medskip

\noindent {\bf C1}: For any $k\geq 0$, there exists $i\in\{1,\ 2,\ 3,\ 4\}$ such that both the $ii$-entries of $P_k$ and $P_{k+1}$ equal $1/2$.

In other words, in case {\bf C1}, any two consecutive node pair selections share a common node. Then  by induction it can be easily proved  that  ${\rm rank}(\Psi_h)=3$ and $\mathscr{C}(\Psi_h)=3$ for all $h\geq 1$.

\medskip

\noindent {\bf C2}: Suppose {\bf C1} does not hold. Then we can find $k\geq0$, and a permutation $(a, b, c, d)$ of $\{1,\ 2,\ 3,\ 4\}$, such that $P_k=M_{ab}$ and $P_{k+1}=M_{cd}$. We let $k_0$ be the smallest $k$ when such disjoint pairs are selected at time $k$ and $k+1$. The following claim holds by induction.

\medskip

{\it Claim.} For any $h\geq k_0+2$, $\Psi_h$ satisfies one of the following three conditions:
\begin{itemize}
\item[1)] $\mathscr{C}(\Psi_h)=1$;

\item [2)] $\mathscr{C}(\Psi_h)=2$, and there is a permutation $(a',\ b',\ c',\ d')$ of $\{1,\ 2,\ 3,\ 4\}$, such that $[{r}_{a'}]^h=[{r}_{b'}]^h$ and $[{r}_{c'}]^h=[{r}_{d'}]^h$;

\item [3)] $\mathscr{C}(\Psi_h)=3$, and there is a permutation $(a',\ b',\ c',\ d')$ of $\{1,\ 2,\ 3,\ 4\}$, such that $[{r}_{a'}]^h=[{r}_{b'}]^h$ and $[{r}_{a'}]^h=\theta(h)[{r}_{c'}]^h+(1-\theta(h))[{r}_{d'}]^h$, $\theta(h)$ can be written as $\theta(h)=\frac{\theta_1(h)}{\theta_2(h)}$, where $\theta_1(h)\in \mathds{Z}$ is odd, $\theta_2(h)\in \mathds{Z}_{> 0}$ is even.
\end{itemize}

\medskip

Therefore, {\bf C1} and {\bf C2} indicate that i) and ii) in the lemma cannot hold simultaneously, which completes the proof. \hfill$\square$

We are now in a place to prove the desired  proposition by reversing the convergence process.

After step $T$, the four row vectors of  $\Psi_T$ have the same value ${\bf{1}}^T/4$. Without loss of generality, we assume $P_{T-1}=M_{12}$. Since $\Psi_T=P_{T-1}\Psi_{T-1}=M_{12}\Psi_{T-1}$, $[{r}_3]^{T}=[{r}_3]^{T-1}$ and $[{r}_4]^{T}=[{r}_4]^{T-1}$. Then,
$$[{r}_3]^{T-1}=[{r}_4]^{T-1}={\bf{1}}^T/4.$$
Moreover, $[{r}_1]^{T-1}$ and $[{r}_2]^{T-1}$ are two other different values with
$
{\bf{1}}^T/4=[{r}_1]^{T}=[{r}_2]^{T}=\frac{[{r}_1]^{T-1}+[{r}_2]^{T-1}}{2}.
$
So it must be that $P_{T-2}=M_{34}$ and $[{r}_3]^{T-2}\neq [{r}_4]^{T-2}$.

Because $P_{T-2}=M_{34}$ and $\Psi_{T-1}=P_{T-2}\Psi_{T-2}=M_{34}\Psi_{T-2}$, the first and second row vectors of $\Psi_{T-1}$ are the same as those of $\Psi_{T-2}$.
Thus, $[{r}_1]^{T-2}\neq [{r}_2]^{T-2}$. Then, $P_{T-3}$ can not be $M_{34}$ or $M_{12}$.

Without loss of generality, we assume $P_{T-3}=M_{13}$.
Thus,
\begin{equation}\label{t-2}
[{r}_3]^{T-2}=[{r}_1]^{T-2}.
\end{equation}

Since $\Psi_{T}=P_{T-1}P_{T-2}\Psi_{T-2}=M_{12}M_{34}\Psi_{T-2}$,
\begin{equation}\label{t}
{\bf{1}}^T/4=\frac{[{r}_1]^{T-2}+[{r}_2]^{T-2}}{2}=\frac{[{r}_3]^{T-2}+[{r}_4]^{T-2}}{2}.
\end{equation}

According to (\ref{t-2}) and (\ref{t}), we conclude $[{r}_4]^{T-2}=[{r}_2]^{T-2}$. Now that $P_{T-3}=M_{13}$, it must be $[{r}_4]^{T-3}=[{r}_2]^{T-3}$.

Since $P_{T-3}=M_{13}$, $P_{T-4}$ cannot be equal to $M_{13}$. On the other hand   $[{r}_1]^{T-3}=[{r}_2]^{T-3}$ if $P_{T-4}=M_{12}$. This implies that $[{r}_1]^{T-3}=[{r}_2]^{T-3}=[{r}_4]^{T-3}$, which is impossible since it contradicts Lemma \ref{no3points}. Following the same argument, $P_{T-4}$ cannot be equal to $M_{14}$, $M_{23}$ or $M_{34}$ as well.
Thus, it leaves the only option that $P_{T-4}=M_{24}$, which completes the proof.

\subsection*{C. Proof of Lemma \ref{lem3}} First of all, it is easy to verify that $\tilde{f}=(\tilde{f}_1,\ \tilde{f}_2,\ ...,\ \tilde{f}_n)$ defined by
\[\tilde{f}_i=
\begin{cases}
\frac{1}{2^{m+1}}, &\ i=1,\ 2,\ ...,\ 2r\\
\frac{1}{2^{m}}, &\ i=2r+1,\ 2r+2,\ ...,\ n
\end{cases}
\]
satisfies that $\sum_{i=1}^{n}\chi_i(\tilde{f}) =mn+2r$.

Next, we show $\sum_{i=1}^{n}\chi_i({f}) \geq mn+2r$ for all $f\in \mathbb{F}$. For simplicity,  define $\chi(f)=\sum_{i=1}^{n}\chi_i({f})$. For any $f\in \mathbb{F}$,  $b_i,i=1,\dots,n$ and $c_i,i=1,\dots,n$ are uniquely determined, we therefore denote them by $b_i(f)$ and $c_i(f)$, respectively, for $i=1,\dots,n$. Denote $\zeta(f)=\max\{c_1(f),\ c_2(f),\ ...,\ c_n(f)\}$. Let $q=(q_1,\ ...,\ q_n)$ be an element in $\mathbb{F}$ satisfying
$$
q\in\arg \min \Big\{\zeta(g): g\in \arg \min_{f\in \mathbb{F}}\chi(f)\Big\}.
$$
The existence of such $q$ is obvious by its definition.

According to the definition of $q$, we have
\[1=\sum_{i=1}^{n}\frac{b_i(q)}{2^{c_i(q)}}.\]
Multiplying  both side of the above equation by $2^{\zeta(q)}$, we  get
\[\sum_{i=1}^{n}2^{\zeta(q)-c_i(q)}b_i(q)=2^{\zeta(q)}.\]
We know immediately that the cardinality of the set $\mathrm{E_q}=\big\{i:c_i(q)=\zeta(q)\big\}$ must be an  even number.

We shall show that $q$ has a similar form as  $\tilde{f}$: $b_i(q)=1$ and $\zeta(q)-c_i(q)\leq 1$ for all $i=1,\dots,n$.  This property is proved by establishing   the following two claims.

\medskip

\noindent {\it Claim 1}. If $c_i(q)=\zeta(q)$, then $b_i(q)=1$.

\medskip
Suppose the claim is not true. Then there exists an index $j\in\mathrm{E_q}$ such that $b_j(q)=2z+1$ for some $z\in \mathbb{Z}^+$. We establish Claim 1 in the following  two cases.

\begin{quote}
\begin{itemize}
\item There is  $k\in \mathrm{E_q}$ such that $c_{k}(q)=\zeta(q)$ and $b_{k}(q)=1$. Define an element $p=(p_1,\ ...,\ p_n)\in \mathbb{F}$ by
$$
p_j=\frac{z}{2^{\zeta(q)-1}};\  p_{k}=\frac{1}{2^{\zeta(q)-1}};\  p_i=q_i, i\notin \{j,k\}.
$$
Now we have $\chi(p)=\chi(q)-1$ since  $\chi_{k}(p)=\chi_{k}(q)-1$, $\chi_{j}(p)=\chi_{j}(q)$ and $\chi_{i}(p)=\chi_{i}(q)$ for all $i\notin \{j,k\}$.
This contradicts  the fact that $\chi(q)=\min_{f\in \mathbb{F}}\chi(f)$.

\item  For all $i\in \mathrm{E_q}$ satisfying $c_i(q)=\zeta(q)$, it holds that $b_i(q)>1$. As mentioned above the cardinality of the set $\mathrm{E_q}$ is an  even number. We denote the number of elements in $\mathrm{E_q}$ as $2s$ with $s\geq 0$. We label these elements as $k_1,\ \dots,\ k_{2s}$. Since $b_{k_i}(q)$ is odd, they can be expressed as $b_{k_i}(q)=2l_i+1$, where $l_i$ is a positive integer, for $i=1,\ ...,\ 2s$.
Define $p_{k_i}=\frac{l_i}{2^{\zeta(q)-1}}$,  $p_{k_{i+s}}=\frac{l_{i+s}+1}{2^{\zeta(q)-1}}, i=1,\ ...,\ s$, and $p_j=q_j$ for all $j\notin\{k_1,\ \dots,\ k_{2s}\}$. Then $p=(p_1\ ...\ p_n)$ defines an element in the set $F$ with $\chi(p)\leq \chi(q)$ and $\zeta(p)<\zeta(q)$. This leads to a contradiction to the choice of $q$ as well.
\end{itemize}
\end{quote}
\medskip

\noindent{\it Claim 2.}  For all $i=1,\dots,n$, $2^{\zeta(q)-c_i(q)}b_i(q)\leq 2$.

\medskip

Suppose it is not true. Then, there exits a $v$ such that $2^{\zeta(q)-c_{v}(q)}b_{v}(q)>2$. As Claim 1 says, if $c_{v}(q)=\zeta(q)$ then $b_{v}(q)=1$. Therefore, $2^{\zeta(q)-c_{v}(q)}b_{v}(q)\geq 4$. Moreover, there are at least two index $k$ and $w$ such that $c_{k}(q)=\zeta(q)$, $c_{w}(q)=\zeta(q)$, $b_{k}(q)=1$ and $b_{w}(q)=1$. We define $p=(p_1,\ ...,\ p_n)$ in that $p_k=\frac{1}{2^{\zeta(q)-1}}$, $p_{w}=\frac{1}{2^{\zeta(q)-1}}$,  $p_{v}=\frac{2^{\zeta(q)-c_{v}(q)}b_{v}(q)-2}{2^{\zeta(q)}}$, and $p_i=q_i$ for any $i\notin\{k,\ v,\ w\}$.  Since $\chi_{k}(p)=\chi_{k}(q)-1$, $\chi_{w}(p)=\chi_{w}(q)-1$, $\chi_{v}(p)\leq \chi_{v}(q)+1$ and $\chi_{i}(p)=\chi_{i}(q)$ for $i\notin\{ k,\ v,\ w\}$, we have $\chi(p) < \chi(q)$, which contradicts the definition of $q$. This proves Claim 2.

From Claim 2, we conclude that $b_i(q)=1$  and $\zeta(q)-c_i(q)\leq 1$ for all $i=1,\dots,n$. Thus, according to the definition of $q$, one has
\begin{equation}\label{200}
1=\sum_{i=1}^{n}\frac{b_i(q)}{2^{c_i(q)}}=\sum_{i=1}^{n}\frac{1}{2^{c_i(q)}}=|\mathrm{E_q}|\frac{1}{2^{\zeta(q)}}+(n-|\mathrm{E_q}|)\frac{1}{2^{\zeta(q)-1}}.
\end{equation}
where $|\mathrm{E_q}|$ is the number of elements in $\mathrm{E_q}$.
Since $n=2^m+r$, we can solve (\ref{200}) and obtain that $|\mathrm{E_q}|=2r$ and $\zeta(q)=m+1$. As a result, $\chi(q)$ can be computed as $mn+2r$, and this concludes the proof.

\subsection*{D. Proof of Lemma \ref{lempartition}}
Denote $\Phi_h:=T_{h-1} \cdots T_0$ for $h=1,2,\dots$. The induced graph of $\Phi_h$, denoted as $\mathscr{G}_{\Phi_h}=(\mathcal{V}, \mathscr{E}_h)$, is defined in that
$\{m,v\}\in \mathscr{E}_h$ if only if $[\Phi_h]_{mv}\neq 0$ for all $m\neq v\in \mathcal{V}$. We first state a few useful properties:
\begin{quote}
\begin{itemize}
\item[P1.]  Each $\Phi_h$ is doubly stochastic  for all $h=1,2,\dots$ since $T_k,k\geq 0$ are doubly stochastic matrices and so are their products.
\item[P2.] For any $h=1,2,\dots$,  we have $\mathscr{G}_{\Phi_h}= \bigcup_{k=0}^{h-1} \mathcal{G}_k$. This point can be easily verified  noticing that all the diagonal elements of each $T_k$ are positive for all $k\geq 0$. As a result, for any $h\geq 1$, there are $\alpha_0,\dots, \alpha_{h-1} \in \mathbb{R}^+$ 
     such that
    \begin{align} \label{201}
    \Phi_h:=\sum_{k=0}^{h-1}\alpha_k T_k +\Big(1- \sum_{k=0}^{h-1}\alpha_k\Big)I.
    \end{align}
    \item[P3.] The number of connected components of the graph $\bigcup_{k=0}^{h-1} \mathcal{G}_k$ is equal to $4^n-{\rm rank}(\Phi_h)$. Based on (i), (ii), this point becomes clear seeing that $I-\Phi_h$ defines a weighted Laplacian of the graph $\bigcup_{k=0}^{h-1} \mathcal{G}_k$ (cf., Lemma 13.1.1 in \cite{GraphTheory}).
\end{itemize}
\end{quote}

We also need the following lemma to complete the proof.

\medskip

\begin{lemma}\label{lemgraph}
If $\bigcup_{k=0}^{h-1} \mathrm{G}_k$ is connected for some $h\geq 1$, then $\bigcup_{k=0}^{h-1} \mathcal{G}_k$ has $\tau_0={\rm dim} \big\{{\rm \bf vec}(z):  \frac{1}{n!} \sum_{\pi \in \mathbf{P}} U_\pi z U_\pi  =z\big\}$ components.
\end{lemma}

{\it Proof.} Take $\rho \in \mathbb{C}^{2^n\times 2^n}$ and let  $\bigcup_{k=0}^{h-1} \mathrm{G}_k$ be connected.  Denote $\bar{\alpha}_k:={\alpha_k}/{ \sum_{t=0}^{h-1}\alpha_t}$ with $\alpha_k$ specified in (\ref{201}). The following equalities hold:
\begin{align} \label{r1}
&\Big\{{\rm \bf vec}(z): \Phi_h {\rm \bf vec}(z) ={\rm \bf vec}(z), z \in\mathbb{C}^{2^n \times 2^n}\Big\}\nonumber \\
&\stackrel{a)}{=} \Big\{{\rm \bf vec}(z): \sum_{k=0}^{h-1}\bar{\alpha}_k T_k  {\rm \bf vec}(z) ={\rm \bf vec}(z), z \in\mathbb{C}^{2^n \times 2^n}\Big\}\nonumber \\
&\stackrel{b)}{=}\Big\{{\rm \bf vec}(z): \sum_{\{j,m\}\in \bigcup_{k=0}^{h-1} \mathrm{G}_k} \bar{\alpha}_k U_{jm}z U_{jm}^\dag=z,z \in\mathbb{C}^{2^n \times 2^n}\Big\}\nonumber\\
  &\stackrel{c)}{=} \Big\{{\rm \bf vec}(z): U_{jm}z U_{jm}^\dag =z, \{j,m\}\in \bigcup_{k=0}^{h-1} \mathrm{G}_k, z \in\mathbb{C}^{2^n \times 2^n}\Big\} \nonumber\\
  &\stackrel{d)}{=} \Big\{{\rm \bf vec}(z): U_\pi z U_\pi^\dag =z, \pi\in \mathbf{P}, z \in\mathbb{C}^{2^n \times 2^n}\Big\} \nonumber\\
  &\stackrel{e)}{=} \Big\{{\rm \bf vec}(z):  \frac{1}{n!} \sum_{\pi \in \mathbf{P}} U_\pi z U_\pi^\dag =z,z \in\mathbb{C}^{2^n \times 2^n}\Big\}.
  \end{align}
Here $a$) holds from (\ref{201}); $b$) is obtained by plugging in the definition of $T_k$;  $c$) is based on   Lemma 5.2 in \cite{PRAinformation}; $d$) is from the fact that the swapping permutations along each edge of a connected graph  consist of a generating set of the group $\mathbf{P}$. The equivalence of $d$) and $e$) is obtained by that
$$
U_\pi z U_\pi^\dag =U_\pi  \Big( \frac{1}{n!} \sum_{\pi' \in \mathbf{P}} U_{\pi'} z U_{\pi'}^\dag\Big)  U_\pi^\dag=\frac{1}{n!} \sum_{\pi \in \mathbf{P}} U_\pi z U_\pi^\dag=z
$$
if $\frac{1}{n!} \sum_{\pi \in \mathbf{P}} U_\pi z U_\pi^\dag=z$ since $\pi \mathbf{P}=\mathbf{P}$ for any $\pi\in \mathbf{P}$.

Note that (\ref{r1}) immediately  implies that
$$
{\rm ker}\big(I-\Phi _h\big):=\big\{{\rm \bf vec}(z):  \frac{1}{n!} \sum_{\pi \in \mathbf{P}} U_\pi z U_\pi  =z, z \in\mathbb{C}^{2^n \times 2^n}\big\},
$$
which in turn  yields that $\bigcup_{k=0}^{h-1} \mathcal{G}_k$ has $\tau_0={\rm dim} \big\{{\rm \bf vec}(z):  \frac{1}{n!} \sum_{\pi \in \mathbf{P}} U_\pi z U_\pi  =z\big\}$ components in light of P3 stated above. This proves the desired lemma. \hfill$\square$

Now that  both $\bigcup_{k=0}^{h-1} \mathrm{G}_k$ and $\bigcup_{k=0}^{h-1} \mathcal{G}_k$ are non-decreasing in $h$, Lemma \ref{lempartition} can be directly concluded from Lemma \ref{lemgraph}.

\medskip

\section*{Acknowledgments}
The authors are grateful to Dr. Daoyi Dong and Prof. Ian R. Petersen, University of New South Wales at Canberra, Australia, for their introduction to quantum mechanics and algorithms as well as for their inspiring discussions. The authors also thank Prof. Alexandre Proutiere, KTH Royal Institute of Technology,  for him suggesting  the possible structures of finite-time convergent algorithms, which eventually motivated  us for proving Proposition \ref{unique}. G.~Shi and K. H. Johansson are supported by Knut and Alice Wallenberg Foundation, the Swedish Research
Council, and KTH SRA TNG. B. Li is supported  by  NKBRPC (2011CB302400),  NSFC of China (11301518), and  the National Center for Mathematics and Interdisciplinary Sciences, CAS.

\end{document}